\documentclass[runningheads]{llncs}
\usepackage{graphicx}
\usepackage{subfigure}
\usepackage{amsmath}
\usepackage{floatrow}

\usepackage{microtype}      
\usepackage{multirow}
\newfloatcommand{capbtabbox}{table}[][\FBwidth]
\begin{document}
%
\title{mr$^2$NST: Multi-Resolution and Multi-Reference Neural Style Transfer for Mammography}
\titlerunning{Mammography Style Transfer}
%
\author{Sheng Wang$^{1,2}$, Jiayu Huo$^{1,2}$, Xi Ouyang$^{1,2}$, Jifei Che$^{2}$, Xuhua Ren$^{1}$, Zhong Xue$^{2}$, Qian Wang$^{1}$, and Jie-Zhi Cheng$^{2}$}


\institute{Institute for Medical Imaging Technology, School of Biomedical Engineering, Shanghai Jiao Tong University, Shanghai, China\\
\email{wsheng@sjtu.edu.cn}\and
Shanghai United Imaging Intelligence Co., Ltd., Shanghai, China}


\maketitle              
\begin{abstract}
Computer-aided diagnosis with deep learning techniques has been shown to be helpful for the diagnosis of the mammography in many clinical studies. However, the image styles of different vendors are very distinctive, and there may exist domain gap among different vendors that could potentially compromise the universal applicability of one deep learning model. In this study, we explicitly address style variety issue with the proposed multi-resolution and multi-reference neural style transfer (mr$^2$NST) network. The mr$^2$NST can normalize the styles from different vendors to the same style baseline with very high resolution. We illustrate that the image quality of the transferred images is comparable to the quality of original images of the target domain (vendor) in terms of NIMA scores. Meanwhile, the mr$^2$NST results are also shown to be helpful for the lesion detection in mammograms. 


\end{abstract} 
\section{Introduction}

\begin{figure}[!t]
\centering
\includegraphics[width=12.2cm, height=2.833cm]{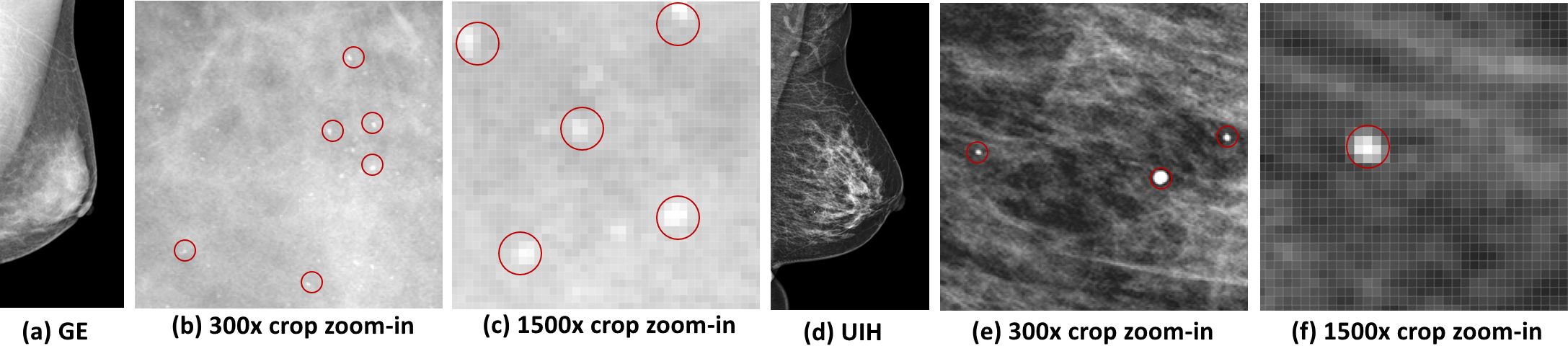}
\caption{The comparison of image styles from different vendors. Red circles highlight calcifications and $\mu$Cs.}
\label{fig:differentstyles}
\end{figure}

The mammography is a widely used screening tool for breast cancer. It has been shown in many studies \cite{freer2001screening}\cite{kooi2017large} that the incorporation of CAD softwares in the reading workflow of mammography can be helpful to improve the diagnostic workup. Equipped with the deep learning (DL) techniques, the CAD scheme was shown to further outperform radiologists from multiple centers across several western countries \cite{screenpoint2019AI}. Although promising CAD performances for mammography have been shown in many previous studies, there is still an essential issue that is not well and explicitly addressed in previous DL works. As shown in Fig. \ref{fig:differentstyles}, the image styles, like image contrast, edge sharpness, etc., of different vendors are quite different. Accordingly, one DL based CAD scheme may not always perform well on mammograms from different vendors, unless sufficient large and diverse training data are provided. Because the collection of large mammograms from various vendors can be very difficult and expensive, we here propose a mammographic style transfer (mST) scheme to normalize the image styles of different vendors to the same style baseline. It will be shown that style normalization step with the mST scheme can further boost the robustness of the classic Faster-RCNN detector \cite{ren2015faster} to the mammograms of different vendors and improve the detection performance for masses and microcalcification, denoted as $\mu$C for short throughout this paper.

The direct use of the off-the-shell neural style transfer (NST) methods for the mST scheme may encounter two major issues. First, the style transfer of very subtle but important abnormalities like $\mu$C or calcification is very challenging. It is because the ST for the $\mu$C, which could be depicted in less than 20 pixels, may need to be carried out in high resolution. However, to our best knowledge, most classic NST methods only support images with resolution less than $1000 \times 1000$, whereas the dimensionality of nowadays mammography is usually larger than $2000 \times 2000$. Therefore, the step of image downsize is inevitable in our problem and hence the quality of subtle abnormalities like $\mu$C after transfer may be compromised. Second, for most classic NST methods, a style reference image is usually needs to be manually selected as network input. However, in our context, an automatic selection scheme for style reference images is needed to facilitate the style normalization process. Meanwhile, the appearance variety of mammography is large and also depends on the category of breast density and the subject's figure. The consideration of only one style reference image may not be sufficient to yield a plausible transfer results. 

To address the two issues, the mST scheme is realized with a new multi-resolution and multi-reference neural style transfer (mr$^2$NST) network in this study. By considering multi-resolution, the details of subtle abnormalities like $\mu$C or calcification can be better preserved in the transfer process. With the multiple reference images, our mr$^2$NST network can deal with wide variety of mammography and integrate the style transfer effects from the reference images for more plausible style normalization results. Our mr$^2$NST network also takes into account the similarities between the input image to be transferred and reference images for the integration of multiple style transfer effects. To our best knowledge, this is the first study that explicitly explores the style transfer technique to mitigate the style variation problem, which may compromise the detection performance for breast lesions.

\begin{figure}[!ht]
\centering
\includegraphics[width=12cm, height=6.8cm]{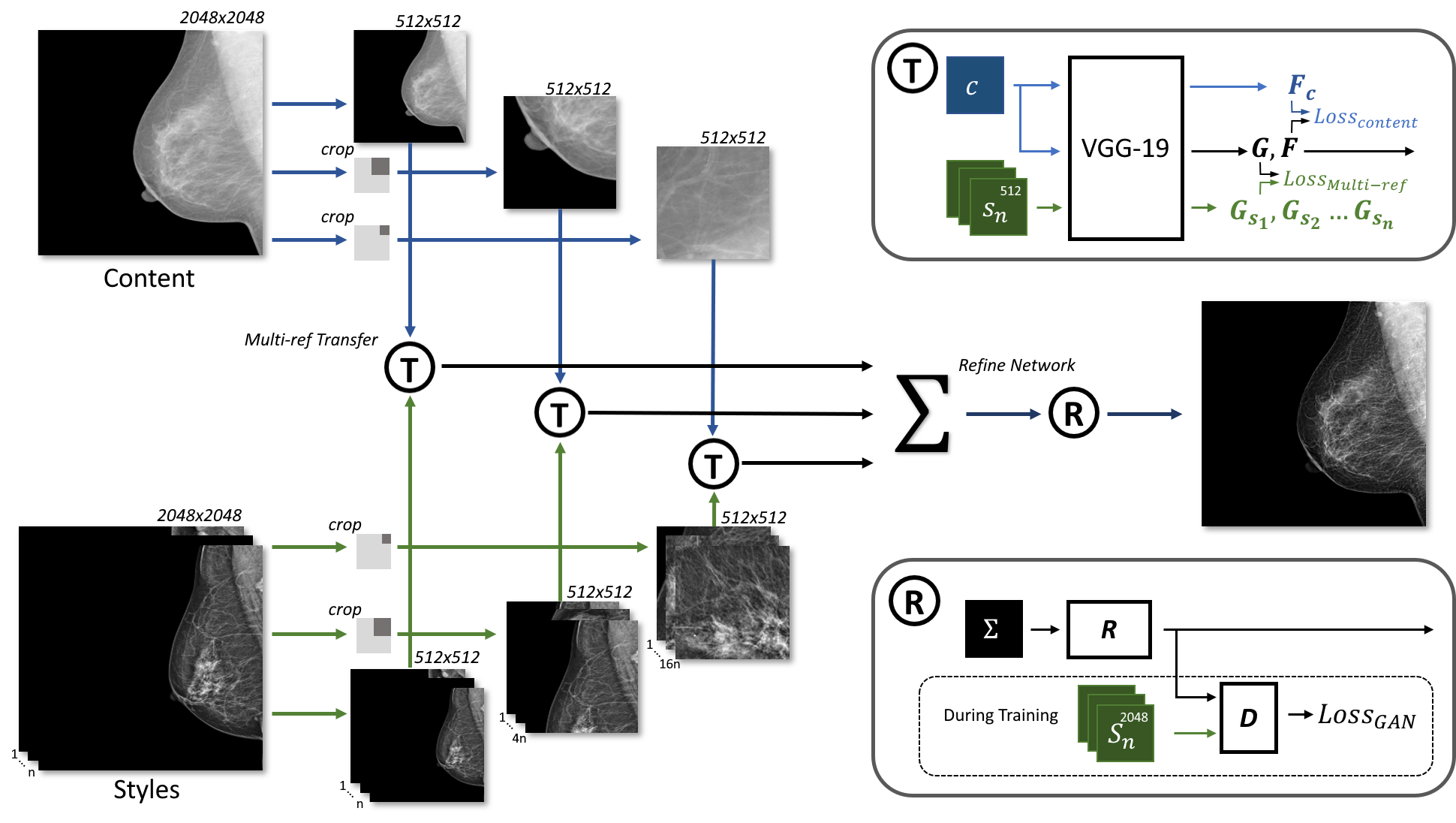}
\caption{The pipeline of the proposed mr$^2$NST for mST. \textbf{T} denote the multi-resolution operation; \textbf{R} stands for the refiner network.}
\label{fig:framework}
\vspace{-.8em}
\end{figure}

We perform the style transfer experiments by comparing with the classic cycleGAN \cite{zhu2017unpaired} and the conventional exact histogram matching (EHM) \cite{coltuc2006exact}, and test the style normalization, i.e., mST, results on the detection tasks of masses and $\mu$Cs in mammograms. The experimental results suggest that the mST results from our mr$^2$NST network are more plausible and can mitigate the problem of style differences from distinctive vendors for better detection results.


\section{Method}

In this section, we will briefly introduce the concept of NST and then discuss the details of our mr$^2$NST network. The network of our mr$^2$NST network is shown in Fig. \ref{fig:framework}, and the backbone is VGG19 \cite{simonyan2014very}.
\subsection{Neural Style Transfer}
The NST, which was first introduced by Gatys et al. \cite{gatys2016image}, commonly requires two input images of a content image $x_C$ to be transferred and a style reference image $x_S$, and then performs feature learning of the feature representatives of $F_l(x_C)$ and $F_l(x_S)$ in layer $l$ of a NST network. 
Each column of $F_l(x)$, 
$F_l \in {R}^{M_l(x) \times N_l}$, is a feature map, 
whereas $N_l$ is the number of feature maps in layer $l$ and $M_l(x)=H_l(x) \times W_l(x)$ 
is the product of height and width of each feature map. 
The output of NST is the style transferred image, denoted as $\hat{x}$, by minimizing the loss function:



\begin{equation}
\label{eq:st_equation}
L_{total} = L_{content} + L_{style},
\end{equation}
where the content term $L_{content}$ compares feature maps from the $\hat{x}$ and $x_S$ of each single layer $l_C$:
\begin{equation}
\label{eq:content_term}
    L_{content} = \dfrac{1}{N_{l_c}M_{l_c}(x_C)} \sum_{ij}{} (F_{l_c}(\hat{x})-F_{l_c}(x_C))^2_{ij},
\end{equation}
and the style term $L_{style}$ compares a set of summary statistics:
\begin{equation}
\label{eq:style_term}
\begin{split}
    L_{style} & = \sum_{l}{} w_l E_l;\\
    E_l & = \dfrac{1}{4N^2_l} \sum_{ij}{} (G_l(\hat{x})-G_l(x_S))^2_{ij},
\end{split}
\end{equation}
where $G_l(x) = \dfrac{1}{M_l(x)}F_l(x)^TF_l(x)$ is the gram matrix of the feature maps of the layer $l$ in response to image $x$.

\subsection{Multiple Reference Style Images}
The mr$^2$NST network takes multiple reference style images for better accommodate the appearance variety of mammography. Different regions in a mammography may need distinctive reference images to be transferred. For example the dense breast image to be transferred may be more suitable to take reference of images with denser glandular tissues. To attain this goal, a quantitative measurement for style similarity is needed.

The gram matrix in the equation \ref{eq:style_term} computes the co-variance statistics of features at one layer as the quantification of style similarity of the corresponding perceptual level. A higher value in the gram matrix suggests more similar of the corresponding paired feature maps in style. Accordingly, with the multiple $n$ reference style images, we can compute the corresponding gram matrices with each style image and integrate of the gram matrices with the max operation. Specifically, A simple but effective multi-reference style term is defined as 

\begin{equation}
\begin{split}
\label{eq:sty_mref}
    L_{Multi-ref} & = \sum_{l}{} w_l E_l;\\
    E_l & = \dfrac{1}{4N^2_l} \sum_{ij}{} (G_l(\hat{x})-G_l)^2_{ij};\\
    G_l & = H(M(F_l(x_{S_1}),F_l(x_{S_2})...F_l(x_{S_n})), \overline{h}).
\end{split}
\end{equation}
The function $M()$ is a element-wise max operation takes $n N_l \times H_l \times W_l$ feature maps $F(x_{S_n})$  with the $n$th reference image at the $l$th layer and outputs a $N\times N$ matrix, $G\sp{\prime} _l$. Specifically, the function $M()$ computes each element $g\sp{\prime} _{ij}$ of $G\sp{\prime} _l$ as

\begin{equation}
    g{\prime}_{ij} = max \Big( F_i(x_{S_1})^T F_j(x_{S_1}),F_i(x_{S_2})^T F_j(x_{S_2}), ..., F_i(x_{S_n})^T F_j(x_{S_n}) \Big).
\end{equation}
The function $H$ is a histogram specification function to normalize the $G\sp{\prime} _l$ with the reference density histogram, $\overline{h}$, for numerical stabilization. $\overline{h}$ is the density histogram of a $n\times N_l \times N_l$ matrix stacked by $n$ $N_l \times N_l$ style gram matrices.

The size of nowadays mammography is commonly bigger than $2080 \times 2080$, and may require formidably large GPU memory for any off-the-shelf NST method. In our experience, the ST for an image with the size of $512 \times 512$ could consume up to 10.8GB GPU memory for inference. For the mST with original size, it is estimated to require more than 160GB GPU memory and hence is very impractical for the clinical usage or laboratorial study. Accordingly, we here propose a multi-resolution strategy that can more efficiently use the GPU resources and still attain the goal of better consideration of local details in the mST scheme. 

Referring to Fig. \ref{fig:framework}, the multi-resolution is implemented by considering the $2048 \times 2048$ original image (scale0), division of image into $4$ $1024 \times 1024$ patches with overlapping (scale1) as well as $16$ $512 \times 512$ patches with overlapping (scale2). The image of scale0 and the patches of scale1 are resized into $512 \times 512$ to fit the memory limit and support the feature learning with the middle- and large-sized receptive fields. 

The image and patches of the scale0, scale1 and scale2 are transferred by taking multiple reference style images, see Fig. \ref{fig:framework}. For each patch/image of each scale, we perform the style transfer by optimizing the multi-reference style term $L_{Multi-ref}$ defined in e.q. \ref{eq:sty_mref}. Afterward, the all transferred patches of scale1 and scale2 are further reconstructed back to the integral mammograms. The reconstructed mammograms of scale1 and scale2 as well as the transferred image of scale0 are then further resized back to the original size. For the final output, we integrate the three transferred images of scale0, scale1 and scale2 with weighted summation and further refine the summed image by a refiner network. The final style transferred mammogram can be computed as
\begin{equation}
    M_{out} = R(S_0,S_1,S_2)=r(\sum_{n=0}^{2}{w_n S_n}),
\end{equation}
where $R$ is the refiner network and $r$ denotes a network composed by $3$ $1 \times 1$ convolutional layers\cite{szegedy2015going}, and $w_0$, $w_1$, and $w_3$ are three learnable weights. The refiner network is trained with the GAN scheme, where the refiner network is treated as generator to fool a discriminator $D$. The discriminator $D$, with the backbone of ResNet18 \cite{he2016deep}, is devised to check whether the input image is of the target style. The training of the refiner GAN can be driven by minimizing the loss:
\begin{equation}
\label{eq:refiner}
    L_{GAN}(R,D) =\log {D(Style)}+\log{(1-D(R(S_0,S_1,S_2)))}.
\end{equation}




\section{Experiments and Results}

\begin{figure}[!ht]
\centering
\includegraphics[width=12.2cm, height=5.86cm]{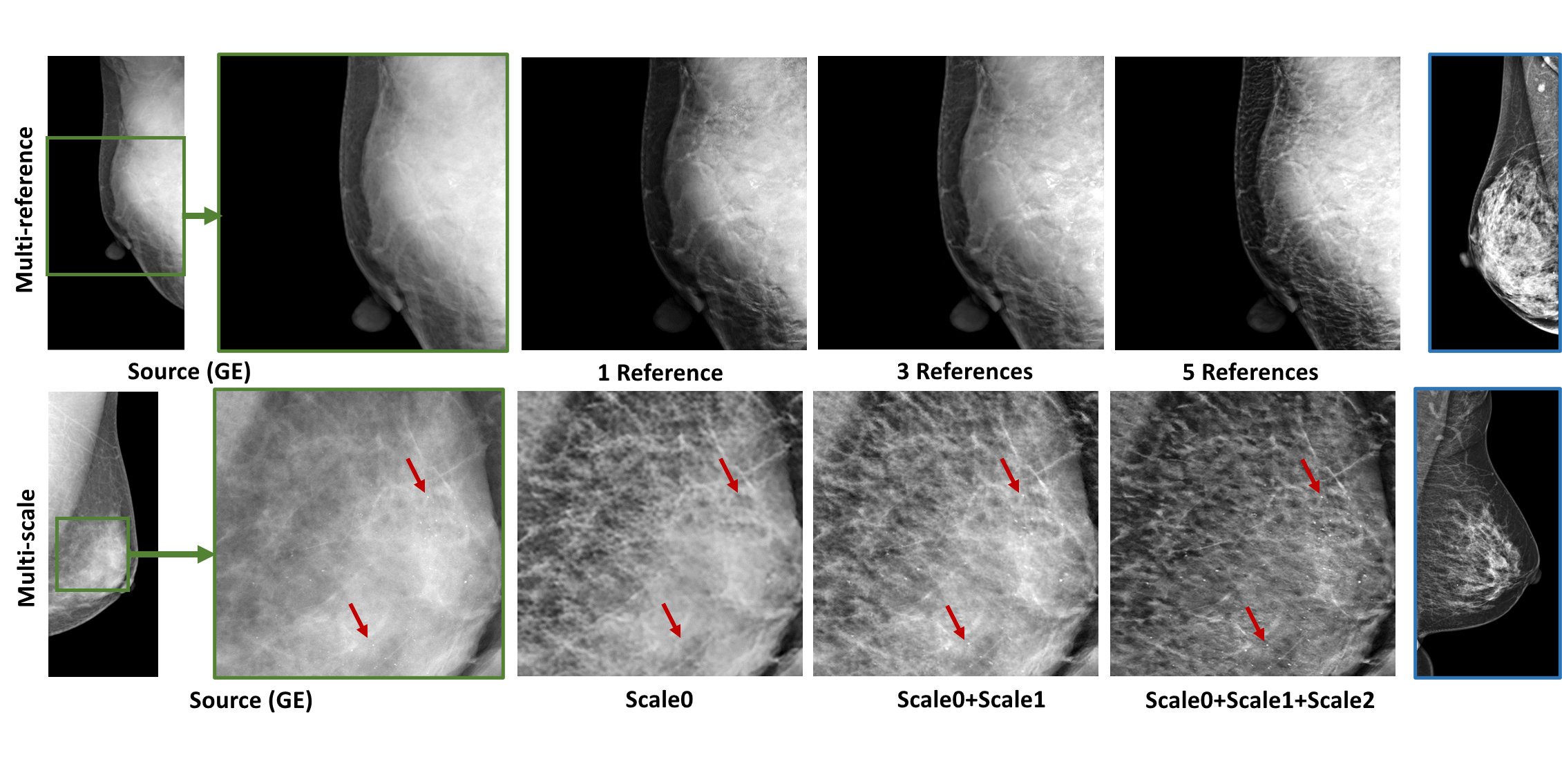}
\caption{The visual illustration of the multiple reference and multiple resolution effect. The red arrows suggest the calcifications or $\mu$Cs in the images.}
\label{fig:multipleres}
\vspace{-.5em}
\end{figure}

\begin{figure}[!ht]
\centering
\includegraphics[width=12.2cm, height=7.51cm]{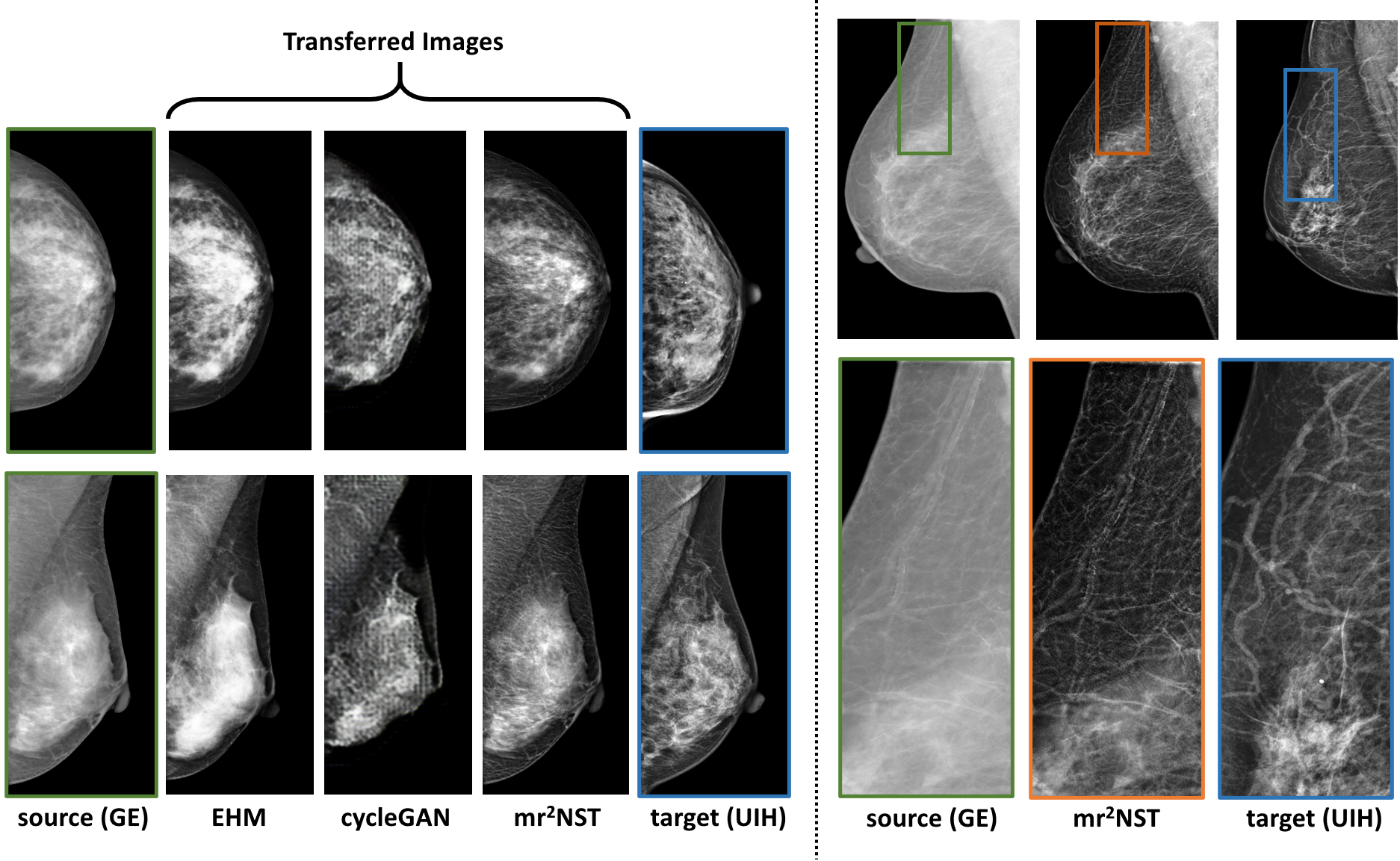}
\caption{Visual comparison of mST results from different methods. The right part gives zoom-in comparison in terms of vessel structures.}
\label{fig:tsne}
\vspace{-.8em}
\end{figure}
In this study, we involved 1,380 mammograms, where 840 and 540 mammograms were collected from two distinctive hospitals, denoted as $H_A$ and $H_B$, with local IRB approvals. The mammograms from $H_A$ and $H_B$ were acquired from the GE healthcare (GE) and United Imaging Healthcare (UIH), respectively. All mammograms are based on the unit of breast. Accordingly, there are half cranicaudal (CC) and half mediolateral oblique (MLO) views of mammograms in our dataset. For the training of the refiner GAN with the e.q. \ref{eq:refiner}, we use independent 80 GE and 80 UIH mammograms, which are not included in the 1,380 images.  




Throughout the experiments, we set the source and target domains as GE ($H_A$) and UIH ($H_B$), respectively. As can be found in Fig. \ref{fig:differentstyles} and Fig. \ref{fig:tsne}, the image style of GE is relatively soft, whereas the UIH style is sharper. Accordingly, the image styles from different vendors can be very distinctive. We compare our method with the baselines of cycleGAN \cite{zhu2017unpaired} and exact histogram matching (EHM)\cite{coltuc2006exact}. Since the cycleGAN requires training step, we randomly select 100 and 80 images from $H_A$ and $H_B$, respectively, to train the cycleGAN. Except the refiner GAN, our mr$^2$NST doesn't need a training step. For each ST inference with mr$^2$NST, we select 5 reference images of the target UIH domain with 5 best similar images from an reference image bank of 40 UIH images, which are not included in the 1,380 images and the 80 training data of refiner GAN. The similarity for the selection is based on the area of breast. The selected reference images are of the same view (CC/MLO) with the source image to be transferred. The optimizer Adam is adopted with 400 epochs of optimization for our mr$^2$NST. 
 
Fig. \ref{fig:multipleres} illustrates the efficacy of multi-reference and multi-resolution scheme for the mST from GE to UIH. The upper row in Fig. \ref{fig:multipleres} shows better enhance on glandular tissues with 5 reference images on a case with high density, while the lower row suggests the calcification can be better enhanced by fusing the transferred images from all three scales. Fig. \ref{fig:tsne} shows the mST results from our mr$^2$NST and the baselines of cycleGAN and EHM. From visual comparison, the quality of the transferred images from mr$^2$NST are much better. The cycleGAN requires large GPU memory and can't support mST in high resolution. Meanwhile, referring to the right part of Fig. \ref{fig:tsne}, mr$^2$NST can preserve the details of vasculature after the mST. 
 
\begin{table}[]
\caption{NIMA scores of UIH, GE and mST results.}
\label{table:nima}
\begin{tabular}{c|c|c|c|c|c}
\hline
           & GE   & UIH   & mr2NST & cycleGAN & EHM \\ \hline
Score & 5.16 $\pm$ 0.12 & 5.43 $\pm$ 0.10 & 5.42 $\pm$ 0.15 & 4.74 $\pm$ 0.22 & 5.29 $\pm$ 0.11 \\ \hline
\end{tabular}
\end{table}
 
Two experiments are conducted to illustrate the efficacy of our mr$^2$NST w.r.t. the transferred image quality and detection performance. The first experiment aims to evaluate the quality of transferred images with the neural image assessment (NIMA) score \cite{talebi2018nima}. Specifically, we randomly select 400 GE and 400 UIH (not overlapped with the training dataset of cycleGAN) for mST. The 400 GE images are transferred to the UIH domain with the comparing methods and the resulted NIMA scores of the transferred images are listed in Table \ref{table:nima}. We also compare the NIMA scores between the transferred and original images at UIH domain with the student t test. The computed $p$-values are $0.58$, $4.76\times10^{-12}$, and $3.34\times10^{-61}$, w.r.t. mr$^2$NST, EHM, and cycleGAN, suggesting that the quality of mST images from mr$^2$NST is not significantly different to the quality of original UIH images in terms of NIMA scores. On the other hand, the quality differences of mST images from EHM, and cycleGAN to the UIH images deem to be significant.


The second experiment aims to illustrate if the mST can help to mitigate the domain gap problem and improve the detection performance. The dataset of UIH ($H_B$) is relatively small, and therefore, we aim to illustrate if the mST from GE to UIH can assist to improve the detection results in the UIH domain. Since the baselines can't yield comparable image quality to the target UIH domain, we only perform this experiments with mr$^2$NST. Specifically, we conduct 5 schemes of various combination of UIH, UIH$^{GE}$ (simulated UIH with mr$^2$NST from GE), and GE data for the training of Faster-RCNN \cite{ren2015faster} with the backbone of resnet50. The detection results for masses and $\mu$Cs are reported in Table \ref{table:detectResUIH}. 

The testing UIH data, which is served as the testing data for all training settings in Table \ref{table:detectResUIH}, include 120 images of 90 positive cases and 30 normal images. The 90 testing positive cases have 36 and 28 images with only masses and $\mu$C, respectively, and 26 images with both. For the training with only real UIH data, there are 420 images with 100 normal cases and 320 positive cases (131 only masses, 123 only $\mu$C, and 66 both). For the 2$^{nd}$ to 5$^{th}$ schemes in Table \ref{table:detectResUIH}, we aim to compare the effects of adding 420 and 840 extra training data with either real GE or UIH$^{GE}$ images. The UIH$^{GE}$ data of the 2$^{nd}$ and 3$^{rd}$ are the mST results from the GE data of 4$^{th}$ and 5$^{th}$ schemes, respectively, and 420 GE images is the subset of 840 GE images. For systematical comparison, the 420 images GE has the same distribution of mass, $\mu$C and normal cases with the real UIH 420 images, whereas the 840 GE images are distributed in the same ratio with double size. 


In Table \ref{table:detectResUIH}, the detection performance is assessed with average precision (AP) and recall with average 0.5 (Recall$^{0.5}$) and 1 (Recall$^{1}$) false-positives (FP) per image. As can be observed, the adding of UIH$^{GE}$ in the training data can better boost the detection performance, by comparing the rows of 2$^{nd}$, 3$^{rd}$ to 1$^{st}$ row in Table \ref{table:detectResUIH}. Referring to 4$^{th}$ and 5$^{th}$ rows in Table \ref{table:detectResUIH}, the direct incorporation of GE data seems to be not helpful for the detection performance. The transferred UIH$^{GE}$ images on the other hand are more similar to the real UIH images and can be served more informative samples for the training of detector.

\begin{table}[]
\label{table:detectResUIH}
\caption{Detection performance comparison.}
\begin{tabular}{c|c|c|c|c|c|c}
\hline
\multirow{2}{*}{Training Data} & \multicolumn{3}{c|}{Masses}           & \multicolumn{3}{c}{$\mu$Cs}          \\ \cline{2-7} 
                               & AP    & Recall$^{0.5}$ & Recall$^{1}$ & AP    & Recall$^{0.5}$ & Recall$^{1}$ \\ \hline
420 real UIH                   & 0.656 & 0.761          & 0.869        & 0.515 & 0.459          & 0.567        \\ \hline
420 real UIH + 420 UIH$^{GE}$  & 0.724 & \textbf{0.823}          & 0.891        & 0.569 & 0.593          & 0.702        \\ \hline
420 real UIH + 840 UIH$^{GE}$  & \textbf{0.738} & 0.811          & \textbf{0.912}        & \textbf{0.670} & \textbf{0.622}          & \textbf{0.784}        \\ \hline
420 real UIH + 420 GE          & 0.641 & 0.741          & 0.847        & 0.555 & 0.509          & 0.651        \\ \hline
420 real UIH + 840 GE          & 0.654 & 0.738          & 0.869        & 0.632 & 0.604          & 0.738        \\ \hline
\end{tabular}
\end{table}

\section{Conclusion}
A new style transfer method, mr$^2$NST, is proposed in this paper to normalize the image styles form different vendors on the same level. The mST results can be attained with high resolution by take multiple reference images from the target domain. The experimental results suggest that style normalization with mr$^2$NST can improve the detection results for masses and $\mu$Cs. 

%
%
\bibliographystyle{splncs04}
\bibliography{egbib}

\end{document}